\documentstyle[11pt, aaspp4,epsf,rotate]{article}

\def\zdot{$\dot{\rm z}$}

\lefthead{B\"ottcher}
\righthead{Coronal cooling and its signatures in the rapid aperiodic variability
of GBHCs}

\begin{document}

\centerline{Accepted for publication in {\it The Astrophysical Journal}, 
vol. 553 (June 1, 2001)}

\title{Coronal cooling and its signatures in the rapid aperiodic variability of 
Galactic Black-Hole Candidates}
\author{M.  B\"ottcher, \altaffilmark{1,2}}
\altaffiltext{1}{Rice University, Physics and Astronomy Department, MS 108\\
6100 S. Main Street, Houston, TX 77005 -- 1892, USA}
\altaffiltext{2}{Chandra Fellow}

\begin{abstract}
The most popular models for the complex phase and time lags in the rapid 
aperiodic variability of Galactic X-ray binaries are based Comptonization 
of soft seed photons in a hot corona, where small-scale flares are induced
by flares of the soft seed photon input (presumably from a cold accretion
disc). However, in their original version, these models have neglected 
the additional cooling of the coronal plasma due to the increased soft 
seed photon input, and assumed a static coronal temperature structure. 
In this paper, our Monte-Carlo/Fokker-Planck code for time-dependent 
radiation transfer and electron energetics is used to simulate the 
self-consistent coronal response to the various flaring scenarios 
that have been suggested to explain phase and time lags observed 
in some Galactic X-ray binaries. It is found that the predictions 
of models involving slab-coronal geometries are drastically different 
from those deduced under the assumption of a static corona. However, 
with the inclusion of coronal cooling they may even be more successful 
than in their original version in explaining some of the observed phase 
and time lag features. The predictions of the model of inward-drifting 
density perturbations in an ADAF-like, two-temperature flow also differ 
from the static-corona case previously investigated, but may be consistent
with the alternating phase lags seen in GRS~1915+105 and XTE~J1550-564. 
Models based on flares of a cool disc around a hot, inner two-temperature 
flow may be ruled out for most objects where significant 
Fourier-frequency-dependent phase and time lags have been 
observed.
\end{abstract}


\section{Introduction}

The X-ray emission from Galactic X-ray binaries is known
to exhibit aperiodic variability on a vast variety of time 
scales, from months down to milliseconds (for reviews see, 
e.g. \cite{vdk95,cui99}). The PCA on board the {\it Rossi 
X-ray Timing Explorer} (RXTE), has accumulated a large amount
of data on the rapid aperiodic variability (RAV) and its 
photon-energy dependence for a large number of X-ray binaries.
This information contains valuable hints about the source of 
high-energy emission in these objects. Since it is generally
believed that the X-ray emission above $\sim 2$ -- 10~keV
is due to Comptonization of soft photons in a hot, tenuous
coronal gas, it is natural to assume that the RAV of this
emission component also reveals detailed information about
the size scales, dominant physical (heating and cooling)
mechanisms, and geometry of the Comptonizing region. 

Theoretical calculations of the expected time-dependent
signatures of Comptonization in a hot, tenuous, static corona had
been done for the case of a homogeneous corona (\cite{mkk88})
and for inhomogeneous temperature and density distributions 
in the corona (Kazanas, Hua \& Titarchuk \markcite{kht97}1997,
Hua, Kazanas \& Titarchuk \markcite{hkt97}1997) for the case of
central injection of soft photons into a spherical region. This 
had been generalized to other geometries, including the case 
of a cool, outer accretion disc adjacent to an inner, hot, 
quasi-spherical two-temperature inflow by B\"ottcher 
\& Liang (\markcite{bl98}1998). While such models, in
particular in the case of central injection, representative
of a slab-coronal geometry, could successfully reproduce 
the dominant spectral and timing features of some objects, 
such as Cyg~X-1, they generally required that the corona
on top of the geometrically thin, optically thick accretion
disk should extend out to radii of $R \gtrsim 10^4 \, R_s$,
where $R_s$ is the Schwarzschild radius of the accreting
compact object, and that out to those radii, the rate of
energy dissipation per unit volume would have to decrease
only $\propto r^{-1}$, implying that most of the energy
is dissipated at large distance from the central object. 
This seems to be very hard to reconcile with realistic 
calculations of the vertical structure of accretion-disk 
coronae (\cite{mmh94,rc00,meyer00b}) and the radial structure 
of optically thin, ADAF-type flows (\cite{manmoto00,meyer00a,meyer00b}), 
which indicate that the energy dissipation is concentrated much more 
strongly towards the disk surface and the central object, respectively.
The required large coronal size scales were implied by 
the fact that, in these Comptonization models, the 
maximum time lag achievable between hard and soft 
X-rays corresponds to the difference in photon diffusion 
time in the process of Compton upscattering and is 
thus of the order of the light travel time through the 
corona. 

In order to avoid this size-scale problem, two alternative models
for the hard time lags in Galactic X-ray binaries have been developed.
In the first one (\cite{bl99}), an inner, hot coronal flow is assumed 
to be surrounded by a cool, outer disc. In irregular intervals, blobs 
of cool material are breaking lose from the cool disc and spiral inward 
through the hot coronal flow. (During the inspiraling process they may 
most certainly be tidally disrupted into ring-like density perturbations, 
but for simplicity, we will continue to call these inward-drifting
perturbations ``blobs''. It has been shown in B\"ottcher \& Liang
[\markcite{bl99}1999] that they maintain their integrity as isolated,
dense regions, and do not evaporate during the inspiraling process). 
In this model, hard lags are due to the spectral hardening resulting
from the inward-movement of the soft photon source into hotter
regions of the coronal flow, combined with the gradual heating
of the blob. Thus, the maximum hard time lag achievable corresponds
to the drift time scale of the blobs, which may be several orders of
magnitude longer than the light travel time through the corona, thus
allowing for a correspondingly smaller corona.

In the second alternative model (\cite{pf99}), hard X-ray flares are
assumed to be produced by magnetic flares in a hot corona on top of
a cool accretion disc. Thus, in this model, the flare is produced by
additional heating of the Comptonizing material rather than a flare
of the soft photon input. The required spectral hardening occurs
naturally during the period of efficient energy release in the corona,
and hard lags could be produced in a model consistent with the 
observed weakness of the Compton reflection component if the active 
regions of the corona are driven away from the disc due to radiation
pressure (\cite{beloborodov99}).

Except for the latter model, in which the coronal heating and cooling
are the essential ingredients for modeling the hard X-ray spectral 
evolution, all the Comptonization-based models mentioned above had
so far been calculated under the assumption of a static corona with
fixed electron temperature, which does not change in response to the
increased soft photon input during flares. Since it is generally believed
that the coronal temperature --- at least in spectral states other 
than the very-low or off state of transient sources --- is 
determined by the energy balance between various heating 
mechanisms and cooling dominated by Compton cooling, this may 
be an unrealistic assumption, as has been pointed out by several authors 
(e.g., \cite{mcp00,mj00}). The coronal response to accretion-disc flares 
in the case of a slab-coronal geometry has recently been investigated 
for two special test cases by Malzac \& Jourdain (\markcite{mj00}2000). 
They found that coronal cooling due to the increased soft photon input 
leads to rapid cooling of the coronal electron population and a 
pivoting of the X-ray spectrum around $\sim 20$~keV for typical 
parameters, with rather small amplitude variability in the 
{\it RXTE} PCA energy range. They conclude that the resulting 
variability patterns are inconsistent with the ones observed in 
most X-ray binaries. 

In this paper, I am presenting a detailed re-evaluation of variability
models based on flaring activity in the soft photon input, taking into
account self-consistently the coronal heating and cooling in response
to the time-varying soft photon input. The models considered here are
the slab-coronal geometry, the hot inner two-temperature flow with 
outer, cool accretion disc, and the inward-drifting blob model. 
The model of dynamical active coronal regions has been 
investigated in detail in Poutanen \& Fabian (\markcite{pf99}1999), 
and has also been considered in Malzac \& Jourdain 
(\markcite{mj00}2000). In Section \ref{code}, a brief
description of the coupled Monte-Carlo/Fokker-Planck code used
for this study is given. In Sections \ref{slab}, \ref{adaf}, and
\ref{blob} the results concerning the slab-coronal geometry,
the hot inner two-temperature flow with outer, cool accretion
disc, and the inward-drifting blob case, respectively, are presented. 
Section \ref{summary} contains a brief summary and conclusions.

\section{\label{code}The Monte-Carlo/Fokker-Planck code}

For our time-dependent simulations of the radiation transfer and
electron heating/cooling, the coupled Monte-Carlo Compton scattering 
and Fokker-Planck electron dynamics code described in detail by
B\"ottcher \& Liang (\markcite{bl01}2000) is used. This code
employs a Monte-Carlo method similar to the large-particle method
(\cite{stern95}) for the photon transport, including all
relevant radiation mechanisms, such as Compton scattering,
cyclotron/synchrotron emission and absorption, bremsstrahlung
emission and absorption, and $\gamma\gamma$ absorption and pair
production and annihilation. The evolution of the electron
population is treated simultaneously through an implicit 
Fokker-Planck scheme which allows for arbitrary thermal 
and/or nonthermal electron distributions. At each time step, 
the radiative energy-loss rates are normalized self-consistently 
to the energy transfer between electrons and photons evaluated 
during the photon-transport calculation in the respective
time step. 

Electron acceleration/heating is possible through Coulomb 
interactions with a pool of background, thermal protons or 
through resonant wave-particle interaction (2$^{nd}$ order
Fermi acceleration) with Alfv\'en and whistler wave turbulence,
as well as through self-absorption processes. This means that
--- unlike in most other simulations of related problems (e.g.,
\cite{stern95}; Li, Kusunose, \& Liang \markcite{lkl96}1996;
Dove, Wilms, \& Begelman \markcite{dove97}1996; \cite{mj00}) 
--- the coronal heating compactness is the self-consistent 
result of specific acceleration processes, and is not a 
pre-specified parameter.

For the different geometrical and physical scenarios treated
below, the corona is generally split up into 10 or 15 vertical or
radial zones within which the physical conditions (electron and 
proton densities, magnetic fields, etc.) are assumed to be homogeneous. 
Unless otherwise specified, a globally unordered magnetic field in 
equipartition with the background thermal proton plasma is assumed.

The properties of all photons escaping the corona toward the observer 
are written into an event file, from which photon-energy dependent 
light curves and snapshot photon spectra at arbitrary time intervals 
can be extracted.

\section{\label{slab}The slab-corona case}

Let us first investigate the coronal response to an 
accretion-disc flare in a slab-coronal geometry (e.g., 
\cite{lp77,bk77,hm91,hm93,dove97}). Such flares could
be produced, e.g., by an unsteady accretion flow through the
inner portion of the disk, or by small-scale magnetic flares 
in the ionized surface layer of the disk. For this case, the
soft photon input from the disc is represented by a thermal
blackbody spectrum of $kT_e = 0.2$~keV, typical of the soft
excess in the X-ray spectra of black-hole X-ray binaries in 
the low/hard state. An accretion disc flare is simulated by
increasing the blackbody temperature to 0.5~keV (i.e. increasing
the soft photon compactness by a factor of 39) over a limited 
time interval $\Delta t_{\rm flare}$. Compton reflection
of coronal radiation impinging onto the disc is taken into
account using the Green's functions of White, Lightman, \&
Zdziarski (\markcite{wlz88}1988) and Lightman \& White 
(\markcite{lw88}1988), and at any given time the accretion 
disc flux is enhanced by the amount of coronal flux absorbed 
within the disc. 

The corona is split up into 10 vertical zones of equal height
$\Delta h$. The total height of the corona is fixed to $h =
10^8$~cm, its Thomson depth is $\tau_{\rm T} = 1$, and the
temperature of the background proton plasma is $kT_p = 100$~MeV. 
Coulomb heating of the electrons is assumed to be the dominant
electron acceleration mechanism. This yields an equilibrium
electron temperature of $kT_e \approx 40$~keV. 

\begin{figure}
\epsfysize=12cm
\center{\epsffile{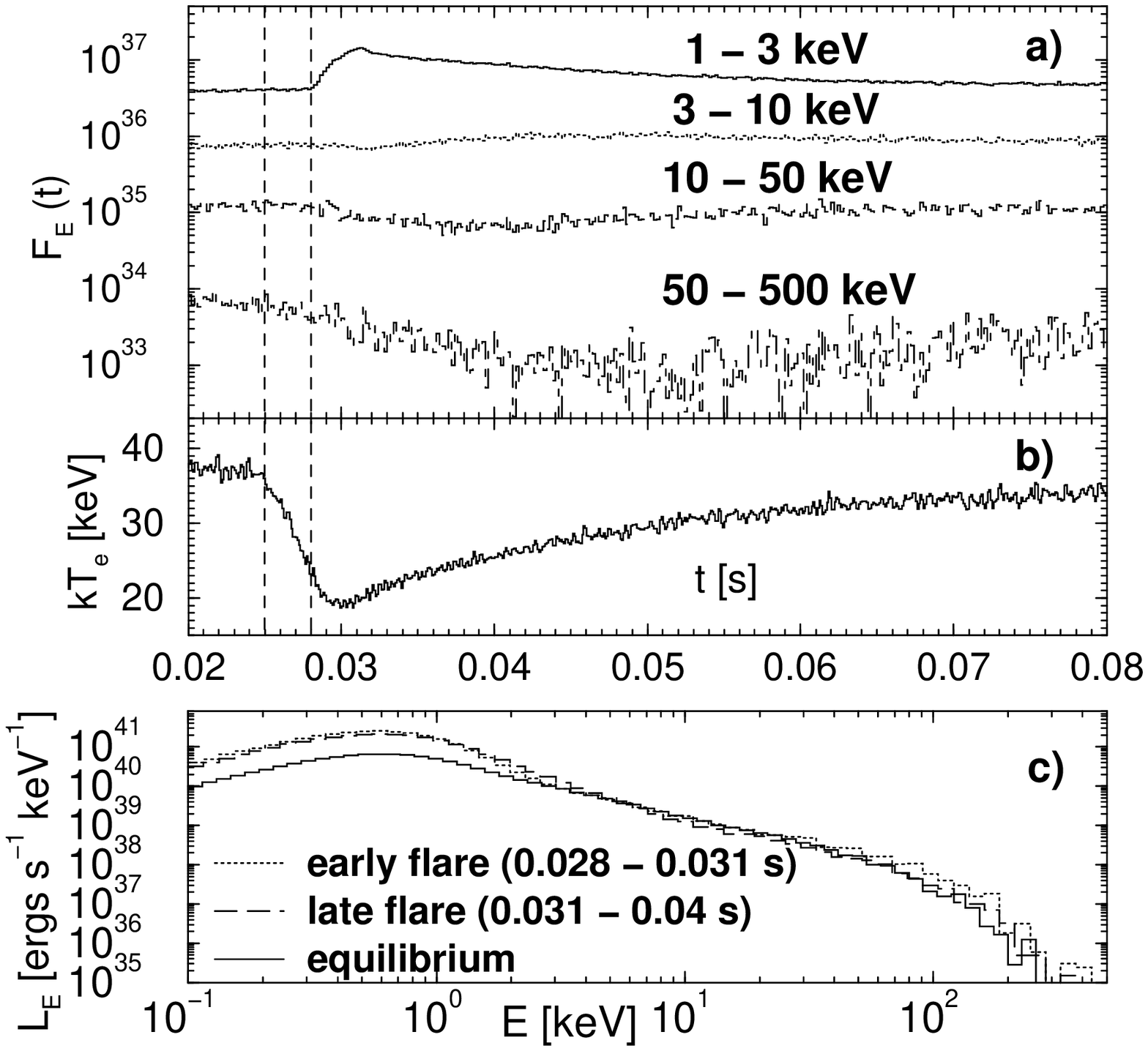}}
\caption[]{Energy-dependent light curves resulting from the coronal
response to an accretion disc flare of duration $\Delta t = 3 \times 
10^{-3}$~s, within which the disc temperature is increased from 0.2 
to 0.5~keV. The vertical dashed lines in panels a) and b) indicate 
the duration of the flare. The corona (slab geometry) has a Thomson 
depth $\tau_{\rm T} = 1$, height $h = 10^8$~cm, and proton temperature 
$kT_p = 100$~MeV. Panel b) shows the average coronal temperature as
a function of time. Panel c) shows snapshot spectra in the early and 
late flare phases and in equilibrium.}
\label{adc1}
\end{figure}

Fig. \ref{adc1} shows the energy-dependent light curves, the
evolution of the average coronal temperature, and some snapshot
energy spectra of the observable X-ray emission for a typical 
simulation, where $\Delta t_{\rm flare} = 3 \times 10^{-3} \,
{\rm s} \approx h/c$. The accretion-disc flare leads to strong
cooling of the coronal electrons. Consequently, in agreement
with the results of Malzac \& Jourdain (\markcite{mj00}2000),
at hard X-ray energies ($E \gtrsim 3$~keV) a broad dip rather 
than a flare results. 

The simulated hard X-ray light curves can be reasonably well
fitted with a function

\begin{equation}
f(t) = \min\left\lbrace F_0 , \; F_1 (t) \right\rbrace,
\label{adc_lightcurves}
\end{equation}
where
\begin{equation}
F_1 (t) = F_1^0 \, \left( e^{- {t - t_0 \over \tau_{\rm d}}} 
\, \Theta [t_0 - t] + e^{t - t_0 \over \tau_{\rm r}}
\, \Theta [t - t_0] \right). 
\end{equation}
Here, $\Theta (x)$ is the Heaviside function.

In a series of simulations, different input parameters, such as the 
coronal Thomson depth, proton temperature, etc. have been varied. 
The most critical parameter in these simulations is the duration 
of the accretion disc flare, $\Delta t_{\rm flare}$, compared to the 
light crossing time through the corona, $h/c$. Table \ref{adc_table}
lists the relevant light curve fitting parameters according to Eq. 
(\ref{adc_lightcurves}) for some of those cases with the standard
coronal parameters listed above. 

\begin{figure}
\epsfysize=12cm
\center{\epsffile{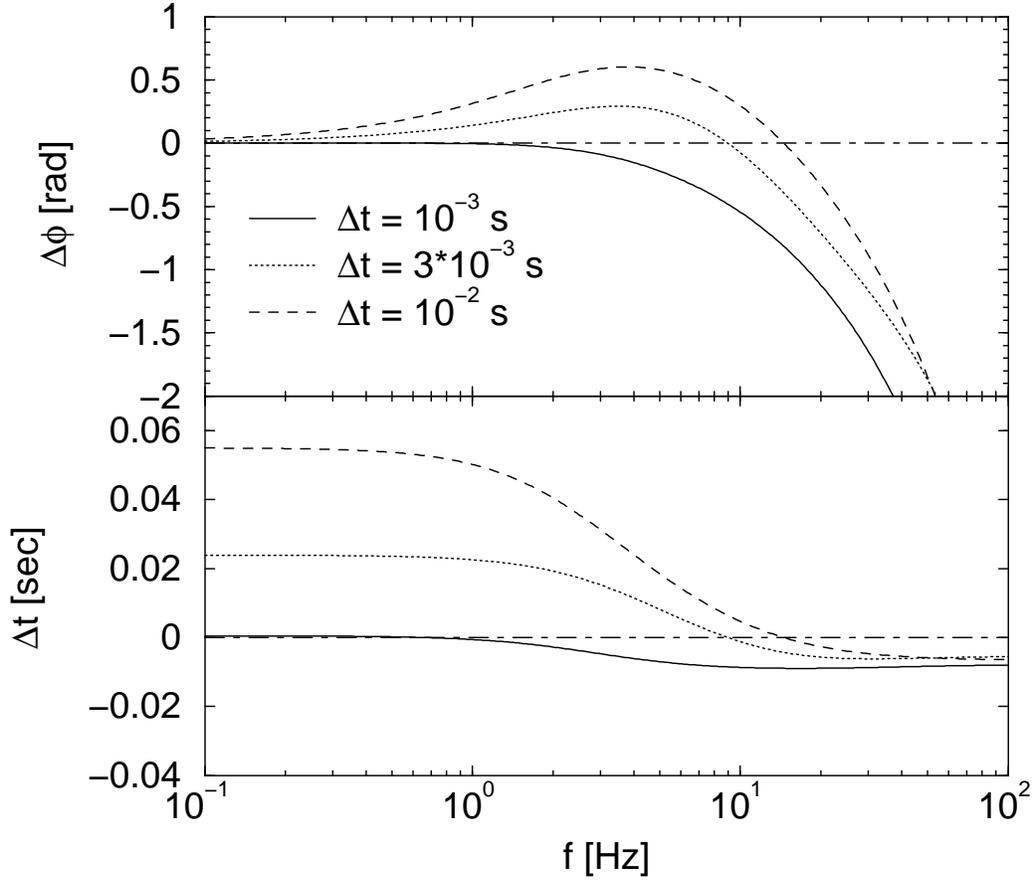}}
\caption[]{Phase lags (upper panel) and time lags (lower panel) 
between the 3 -- 10~keV and the 10 -- 50~keV energy bands
resulting from the coronal response to accretion disc flares of 
different durations in the case of slab-coronal geometry. The 
corona has a Thomson depth $\tau_{\rm T} = 1$, height $h = 10^8$~cm, 
and proton temperature $kT_p = 100$~MeV.}
\label{adc_phaselags}
\end{figure}

\begin{figure}
\epsfysize=12cm
\center{\epsffile{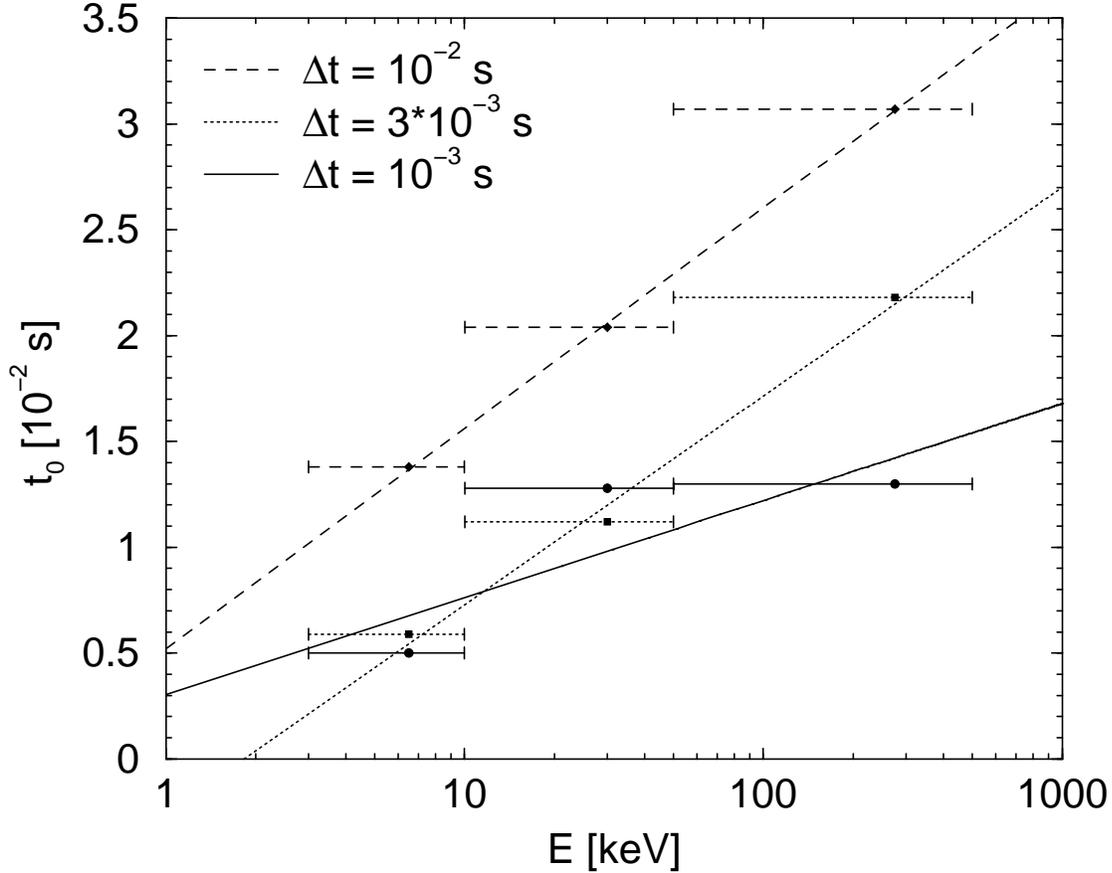}}
\caption[]{Energy dependence of the light-curve fitting parameter
$t_0$, marking the time of the light curve dip, resulting from the 
coronal response to accretion disc flares of different durations in 
the case of slab-coronal geometry. The corona has a Thomson depth 
$\tau_{\rm T} = 1$, height $h = 10^8$~cm, and proton temperature 
$kT_p = 100$~MeV. The curves are best fits of a logarithmic
depencence, $t_0 = A + B \ln(E)$. }
\label{t0_energy}
\end{figure}

The Fourier transformations of the light curves defined in Eq. 
(\ref{adc_lightcurves}) are straightforward, but somewhat lengthy.
If one neglects terms of order $F_1^0/F_0 \ll 1$, the complex
phase of the Fourier transform is given by

\begin{equation}
\psi(\omega) = \phi(\omega) - \omega t_0,
\label{psi}
\end{equation}
where

\begin{equation}
\tan\phi(\omega) = {\omega \, (\tau_{\rm r}^2 - \tau_{\rm d}^2) 
\over (\tau_{\rm r} + \tau_{\rm d}) \, (1 + \omega^2 \, \tau_{\rm r} 
\, \tau_{\rm d})}.
\end{equation}
The resulting phase and time lags between the 3 -- 10~keV and the
10 -- 50~keV bands from the three representative simulations with 
$\Delta t_{\rm flare} = 10^{-2}$~s, $3 \times 10^{-3}$~s, and 
$10^{-3}$~s, respectively, are shown in Fig. \ref{adc_phaselags}. 
The figure illustrates that a rather rich variety of phase and
time lag phenomena can result from this accretion-disc flare 
scenario with a slab-coronal geometry. It appears to be
consistent with a Fourier-frequency independent time lag
at low frequencies, breaking into a power-law with $\Delta t
\propto f^{-\alpha}$, where generally $0.5 \lesssim \alpha
\lesssim 1$ (e.g., \cite{cui97,crary98,nowak99} for Cyg~X-1). 
In the limit $f \ll \tau_{\rm r}^{-1}, \tau_{\rm d}^{-1}$, 
the resulting time lags are dominated by the difference in the
fit parameter $t_0$ for the different energy bands, which
exhibits a photon energy dependence consistent with
the logarithmic time lag dependence measured, e.g., for
Cyg~X-1, GRS~1915+105, (e.g., \cite{cui99}) and XTE~J1550-564
(Wijnands, Homan, \& van der Klis \markcite{wijnands99}1999;
Cui, Zhang, \& Chen \markcite{cui00}2000). The photon-energy
dependence of the parameter $t_0$ is illustrated in Fig.
\ref{t0_energy}. 

Note that this logarithmic energy dependence of the maximum time
lag is related to the time scales for Compton cooling and relaxation 
back to thermal equilibrium due to Coulomb heating rather than 
due to the difference in photon diffusion time between hard and
soft X-rays as in the case of a static corona. This naturally
avoids the problem of the size-scale constraint for static-corona
models for the phase and time lags in the RAV of X-ray binaries
and, instead, allows to place constraints on the proton temperature
and density within the corona, which determine the Coulomb heating
time scale. Assuming that both the electron and proton temperatures
are non-relativistic, the Coulomb heating time scale (which is
expected to be comparable to the maximum time lag between soft
and hard X-rays) can be estimated as

\begin{equation}
\tau_{\rm Coulomb} \sim 3 \times 10^{-3} \, n_{15}^{-1}
{\Theta_e \over \Theta_p} \, (\Theta_e + \Theta_p)^{3/2} \;
{\rm s}.
\label{tau_coulomb}
\end{equation}
(\cite{dl89}). Here, $n_{15}$ is the coronal proton density
in units of $10^{15}$~cm$^{-3}$, and $\Theta_{e,p} = kT_{e,p}
/ (m_{e,p} c^2)$ is the dimensionless electron and proton 
temperature, respectively. Eq. (\ref{tau_coulomb}) indicates
that maximum time lags of order $\sim 10^{-2}$ -- $10^{-1}$~s 
can naturally occur in such a scenario.

Interestingly, Fig. \ref{adc_phaselags} also indicates that
over a limited frequency range also negative phase lags (i.e. 
soft lags) may occur as a result of the coronal response to
accretion-disc flares. Whether this is a potential explanation
for the negative and alternating phase lags observed in
GRS~1915+105 (\cite{cui99b,reig00,lin00}) and XTE~J1550-564 
(\cite{wijnands99,cui00}) remains to be investigated in more
detail in future work.

An important point is that in this scenario, the onset of an
accretion-disc flare marks the onset of the episode of enhanced 
coronal cooling and thus the onset of the decay of the high-energy 
light curves. Consequently, considering that in a realistic scenario
there will be a rapid succession of such flares occurring throughout
the disc, the hard X-ray light curves will exhibit maxima around the 
onsets of the accretion disc flares, with no appreciable offset 
between the maxima at different photon energies. This is consistent 
with the recent result of Maccarone et al. (\markcite{mcp00}2000) 
that the peaks of the cross-correlation function between light curves 
at different energy bands observed in Cyg~X-1 are consistent with
0 time lag. This would be inconsistent with accretion-disc flaring
scenarios in slab-coronal geometry if the observed phase lags were 
due to the time-dependent Comptonization response in a static corona
(\cite{mcp00}).

\section{\label{adaf}The outer-cool-disc case}

An alternative accretion-flow geometry consists of a cool
Shakura-Sunyaev (\markcite{ss73}1973) type, thin disc in the
outer regions of the flow, which exhibits a transition to an
inner, quasi-spherical two-temperature flow, which produces
hard X-rays via Compton upscattering of self-generated synchrotron
and bremsstrahlung photons as well as external soft photons from
the surrounding cool accretion disc. The inner coronal flow could
plausibly be a Shapiro-Lightman-Eardley (\markcite{sle76}1976) type
two-temperature flow, an advection-dominated accretion flow (ADAF; 
\cite{ny94,chen95}) or the recently discovered solution of a 
convection-dominated flow (\cite{qg00}). 

The timing properties which are subject of this paper, are rather 
insensitive to the details of the proton density and temperature 
profile within the inner coronal region. Thus, for simplicity, 
the density and temperature profiles of the inner corona are
fixed corresponding to an ADAF, choosing $n_p \propto r^{-3/2}$, 
and the proton temperature, $T_p \propto r^{-1}$, equal to the
virial temperature. Such a two-phase accretion flow model has
been suggested to explain the various luminosity states of
X-ray binaries (Narayan, McClintock, \& Yi \markcite{narayan96}1996;
\cite{esin98}), and has been very successful, in particular, 
to represent the X-ray spectrum of GRS~1915+105 over a wide 
range of source luminosities (e.g., Muno, Morgan, \& Remillard
\markcite{mmr99}1999). Thus, for a parameter study, I start out from
parameters in the range typically appropriate for spectral fits to 
GRS~1915+105. The inner disc temperature found in GRS~1915+105
is typically 0.75~keV~$\lesssim kT_{\rm BB} \lesssim$~2~keV, and
the inner disc radius is 10~km~$\lesssim r_{\rm in} \lesssim$~100~km
(\cite{mmr99}). As default settings, for simplicity, the soft 
radiation input from the cool, outer disc is represented by a 
thermal blackbody with $kT_{\rm BB} = 1$~keV, and the transition 
radius between cool, outer disc and inner, hot, quasi-spherical flow 
is chosen as $r_{\rm in} = 60$~km. The Thomson depth of the inner
corona is fixed to $\tau_{\rm T} = 0.75$. The event horizon is
modelled as an absorbing inner boundary of the corona at
$r_{\rm horizon} = 10$~km. 

Independent of the value of the transition radius, one finds that
a fraction $f_c = {1 \over 2} \left(1 - {\pi \over 4}\right)$ of 
the accretion disc radiation intercepts the inner corona, if any 
point on the disc surface is assumed to emit isotropically, the 
corona is assumed to be spherical, and the disc has a $T(r) 
\propto r^{-3/4}$ temperature profile. A short accretion-disc 
flare is represented by a temporary increase of the disc blackbody 
temperature to twice its value during non-flaring episodes (i.e. 
increasing the soft photon compactness by a factor of 16). 

\begin{figure}
\epsfysize=12cm
\center{\epsffile{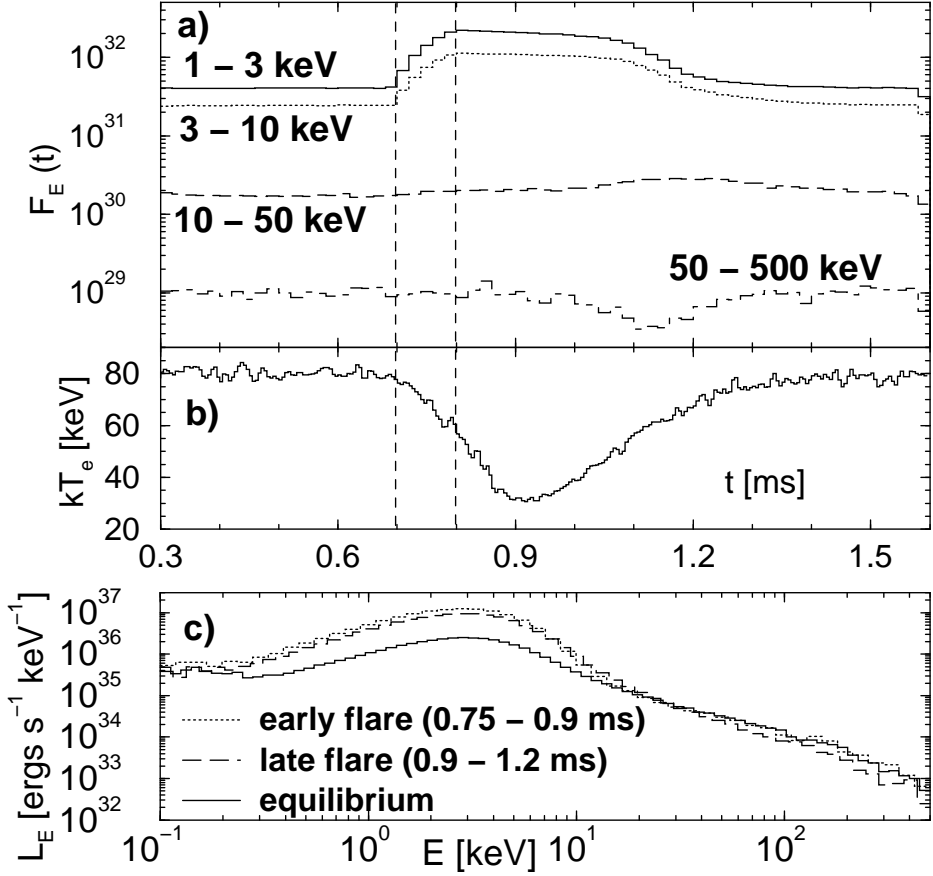}}
\caption[]{Energy-dependent light curves resulting from the response 
of an inner, hot corona to an outer-accretion-disc flare of duration 
$\Delta t = 10^{-4}$~s, within which the disc temperature is 
increased from 1 to 2~keV. The vertical dashed lines in panels a) and 
b) indicate the duration of the flare. The ADAF-like corona has a 
Thomson depth $\tau_{\rm T} = 0.75$, and transition radius (inner disc 
radius) $r_{\rm in} = 60$~km ($r_{\rm in} / c = 2 \times 10^{-4}$~s).
Panel b) shows the average coronal temperature as a function of time. 
Panel c) shows snapshot spectra in the early and late flare phases
and in equilibrium.}
\label{adaf1}
\end{figure}

\begin{figure}
\epsfysize=12cm
\center{\epsffile{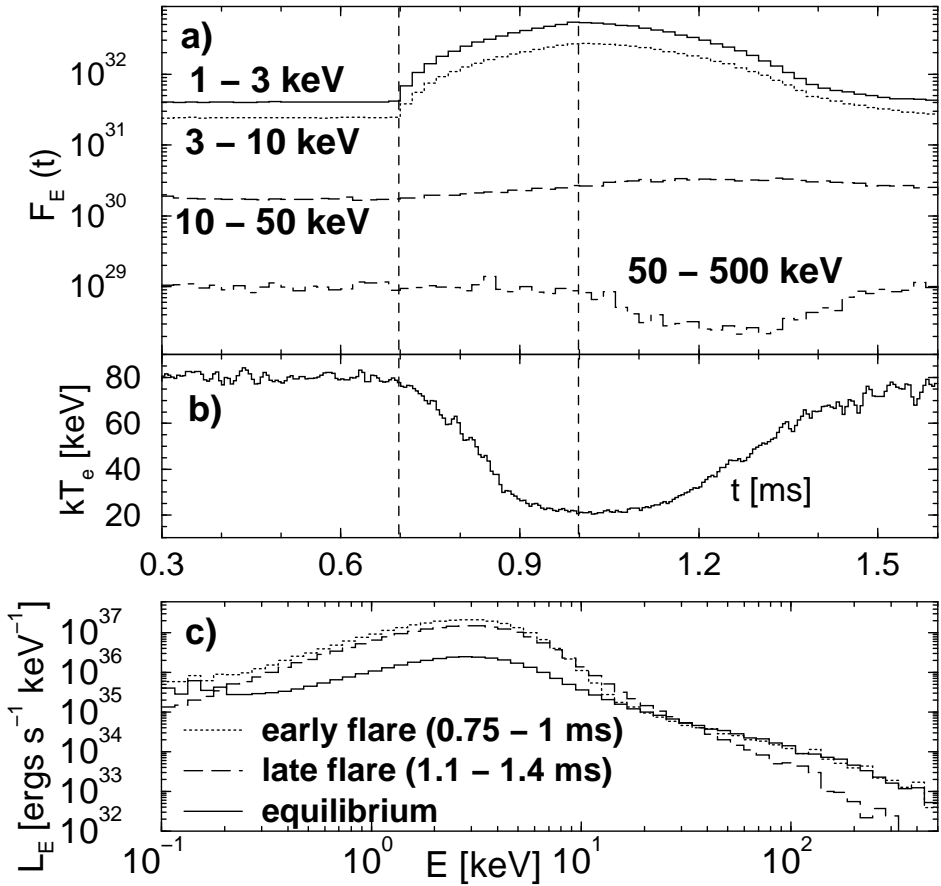}}
\caption[]{Same as Fig. \ref{adaf1}, but with $\Delta t = 3 \times 
10^{-4}$~s.}
\label{adaf2}
\end{figure}

Figs. \ref{adaf1} and \ref{adaf2} shows the energy-dependent light 
curves, evolution of the average coronal temperature, and some snapshot 
energy spectra from two simulations with standard parameters as quoted 
above, and $\Delta t_{\rm flare} = 10^{-4}$~s ($= 0.5 \times r_{\rm in}
/ c$) and $\Delta t_{\rm flare} = 3 \times 10^{-4}$~s ($= 1.5 \times 
r_{\rm in} / c$), respectively. The figures illustrate that in this 
case, the coronal response to the accretion-disc flares occurs on
time scales at least an order of magnitude shorter than in the case
of the slab geometry, due to the substantially higher proton densities 
in the hottest, central regions of the ADAF. This, combined with the 
short light crossing and photon diffusion times through the corona, 
$\sim 10^{-4}$~s, makes it very unlikely that substantial phase and
time lags (of order $\gtrsim 10^{-2}$~s) can be produced by the thermal 
and Comptonization response in such a geometry. In particular, comparing 
the medium and hard X-ray light curves in Figs. \ref{adaf1} and
\ref{adaf2} one finds that if (e.g. in the case of a larger corona)
substantial time lags were produced in this geometry due to radiation
transfer effects in the corona, they would be accompanied by substantial
peak misalignments of the light curves in different energy bands, which
is generally not observed (e.g. \cite{paul98a,paul98b} for GRS~1915+105;
\cite{mcp00} for Cyg~X-1).

\section{\label{blob}The inward-drifting blob case}

The third flaring scenario investigated here is the model of inward 
drifting density perturbations (``blobs'') in a hot, quasi-spherical
inner accretion flow, proposed by B\"ottcher \& Liang 
(\markcite{bl99}1999). This model is based on the accretion-flow
geometry of an outer, cool, optically thick accretion disc and an
inner, hot, quasi-spherical two-temperature flow, as in the previous
section. However, in this case, flaring activity is not dominated
by flares of the outer accretion disc, but by the inward-drifting 
of individual blobs (or ring-like density perturbations) from the
inner boundary of the cool disc through the hot corona, which will
eventually disappear within the event horizon of the black hole.
The spectral hardening required to produce hard phase and time lags
is a consequence of the combined effects of the soft photon source
(the blob) moving further inward toward the hotter and denser
portions of the corona, and of the heating of the blob as it
drifts inward. 

By default, the system of the outer cool disc and the central corona 
will be described by the same set of standard parameters, appropriate
to reproduce typical X-ray spectra of GRS~1915+105 in its high state,
as in the previous section: $\tau_{\rm T} = 0.75$, $n_p \propto r^{-3/2}$, 
virial proton temperature, $r_{\rm in, disc} = 60$~km, $kT_{\rm BB} 
(r_{\rm in}) = 1.0$~keV. It is assumed that the blob surface starts 
out with the same temperature as the disc material, $T_{\rm blob} 
(r_{\rm in}) = T_{\rm BB} (r_{\rm in})$, and heats up as $T_{\rm 
blob} \propto r^{-p}$, where I investigate the special cases 
$p = 0$, $0.25$, $0.5$, and $0.75$. The blob drifts 
inward with a radial drift velocity $\beta_{\rm r} c$. In
this general study, I choose $\beta_{\rm r} = 0.05$. The real 
value of $\beta_{\rm r}$ may in fact be somewhat lower. However,
I choose a rather high radial drift velocity in order to keep the 
required simulation time at $\lesssim 10$~hr (on a 667~MHz Alpha 
Workstation), since the code automatically adjusts the intrinsic 
time step of the simulation to a step size smaller than the time
scale of the net physical heating/cooling rate. For lower values of
$\beta_r$ the resulting timing behaviour is simply stretched over 
a correspondingly longer time interval, and the respective features
in the Fourier power and phase lag spectra below $\sim 1$~kHz (see
below) are shifted to lower frequencies by the same factor, thus
increasing the resulting time lags accordingly (since $\Delta t
= \Delta\phi/[2 \pi f]$).

\begin{figure}
\epsfysize=12cm
\center{\epsffile{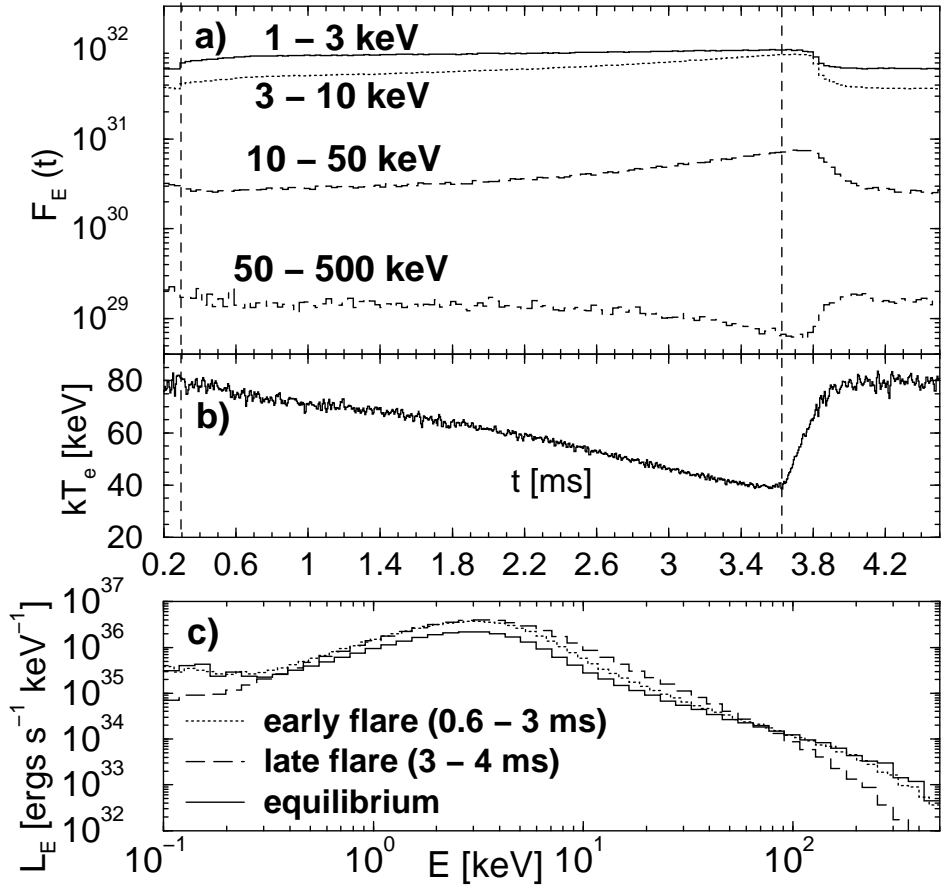}}
\caption[]{Energy-dependent light curves resulting from the 
inward-drifting blob model with coronal cooling. The vertical 
dashed lines in panels a) and b) indicate the time between the
release of the blob at the inner disc radius and the 
disappearence of the blob at the event horizon. The blob
drifts with a radial velocity of 0.05~c and its surface heats
up as $T_{\rm blob} \propto r^{-0.25}$. The ADAF-like corona has a 
Thomson depth $\tau_{\rm T} = 0.75$, and transition radius (inner disc 
radius) $r_{\rm in} = 60$~km ($r_{\rm in} / c = 2 \times 10^{-4}$~s).
Panel b) shows the average coronal temperature as a function of time. 
Panel c) shows snapshot spectra in the early and late flare phases
and in equilibrium. }
\label{blob1}
\end{figure}

A typical simulation result is illustrated in Fig. \ref{blob1}
for a value of $p = 0.25$ (i. e. $T_{\rm blob} \propto r^{-0.25}$). 
Except for the highest energy band (which lies beyond the 
RXTE PCA energy range), the light curves generally show
a gradual rise up to a peak, corresponding to the disappearence of
the blob, and a steep decline. Since this decline happens on sub-ms
time scales, the rapid variability at frequencies $\lesssim 1$~kHz
will be dominated by the gradual-rise phase, which can be parametrized
by a functional form

\begin{equation}
f(t) = F_0 + F_1 \, e^{\left({t - t_0 \over \sigma}\right)^{\alpha}}
\, \Theta(t - t_0) \, \Theta(t_1 - t).
\label{blob_lc}
\end{equation}
The critical fit parameters $\sigma$ and $\alpha$ for various values
of $p$ are listed in Table \ref{blob_table}. Generally, the value
of $\alpha$ increases with increasing photon energy. This increase
becomes more pronounced for decreasing values of $p$. $\sigma$
decreases with increasing photon energy. 

\begin{figure}
\epsfysize=12cm
\center{\epsffile{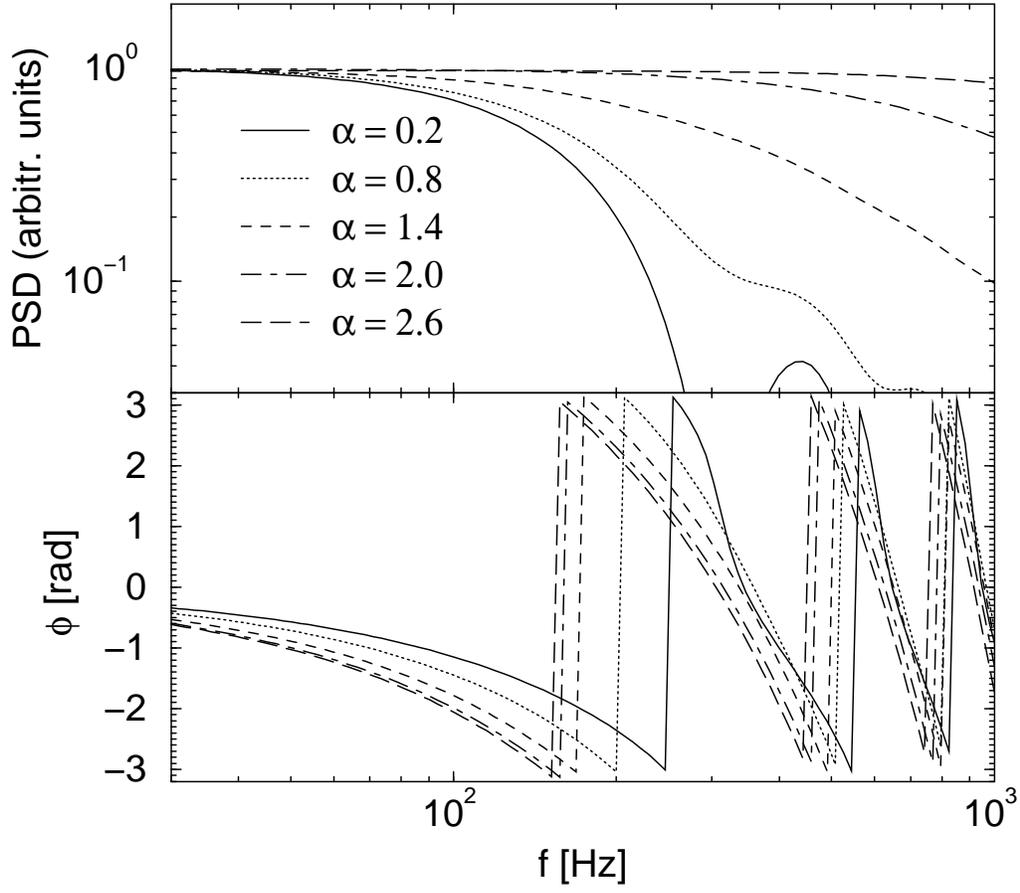}}
\caption[]{Fourier power spectra (upper panel) and Fourier phases
(lower panel) of light curves parametrized by Eq. (\ref{blob_lc})
for various values of $\alpha$. $\sigma = 1$~ms has been chosen.}
\label{blob_fourier}
\end{figure}

For the Fourier transform of Eq. (\ref{blob_lc}), I did not find
an analytical solution. Fig. \ref{blob_fourier} shows the numerically
calculated values of the complex Fourier phase and the power spectra
from light curves parametrized by Eq. (\ref{blob_lc}), for various
values of $\alpha$. $\sigma = 1$~ms has been chosen. The effect of
decreasing $\sigma$ by amounts typical of the variations between
different energy channels (see Table \ref{blob_table}) is similar to,
but much less pronounced than the effect of increasing $\alpha$. 
Thus, the photon-energy dependent power spectral and phase lag 
features are dominated by variations in $\alpha$ between different
energy channels. The figure demonstrates that the inward-drifting
blob model predicts (a) a very significant hardening of the power 
spectra at high frequencies with increasing photon energy and 
(b) a photon-energy and Fourier-frequency dependent phase lag.

\begin{figure}
\epsfysize=12cm
\center{\epsffile{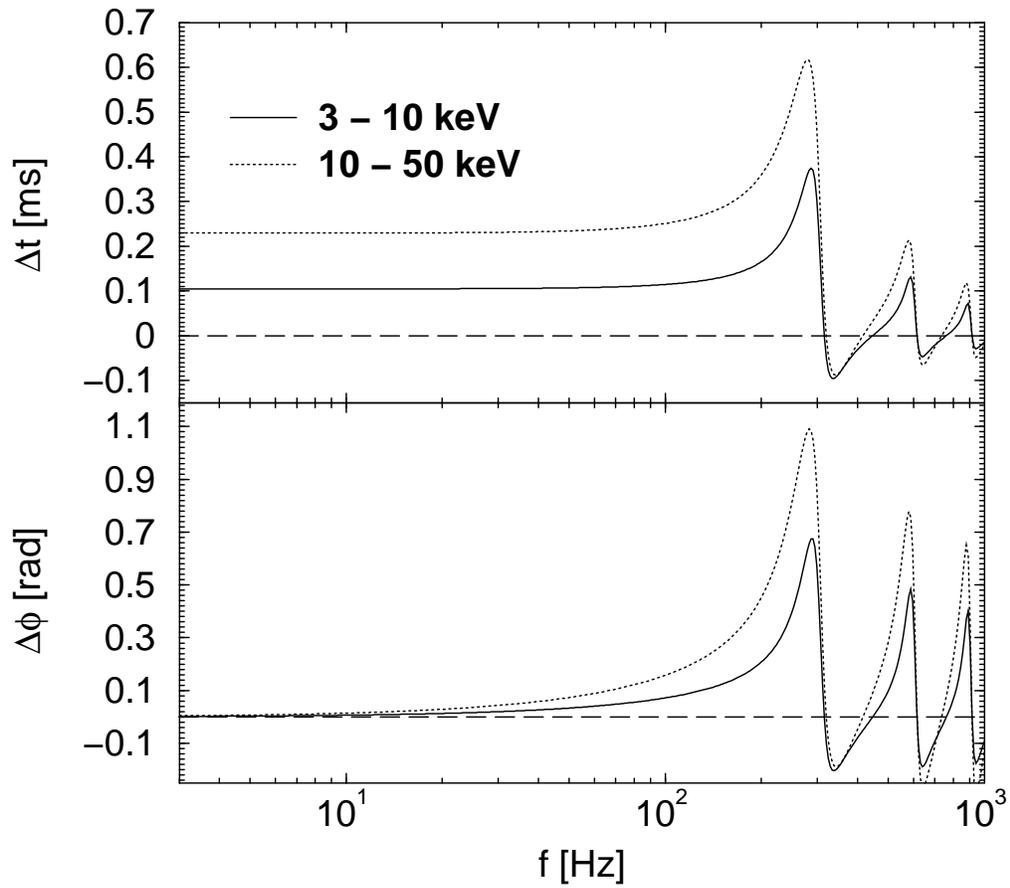}}
\caption[]{Time lag (upper panel) and phase lag (lower panel)
spectra w.r.t. the 1 -- 3~keV channel for the simulation 
illustrated in in Fig. \ref{blob1}. }
\label{blob_phaselags}
\end{figure}

Fig. \ref{blob_phaselags} shows the phase and time lags calculated
from the analytical parametrizations to the simulated light curves
shown in Fig. \ref{blob1}. One can see that --- at least in the
parameter range investigated here ---  the Fourier-frequency dependent
phase lag does generally not show a simple broken-power-law or 
double-broken-power-law behaviour as seen, e.g., in Cyg~X-1 
(e.g., \cite{cui97,crary98,nowak99}). Rather, this scenario 
leads to approximately constant time lags at low Fourier 
frequencies and pronounced positive-lag features around 
the inverse of the blob drift time scale, $f_0  \sim 
(\beta_{\rm r} c) / r_{\rm in}$, beyond which the lags turn
negative in a small frequency interval, and begin to oscillate
between negative and positive values. Beyond $\sim 1$~kHz,
the inadequacy of our analytical description of the light 
curves (Eq. \ref{blob_lc}) on very short time scales ($\lesssim
0.1$~ms) might begin to influence the results, so that no 
predictions can be made at this point. It might be worth to
point out again that all these typical frequencies of the 
features described above scale $\propto \beta_{\rm r}$ and
will thus be shifted towards lower frequencies for lower
values of the drift velocity.

It is conceivable that the oscillating phase lag features
seen in Fig. \ref{blob_phaselags} could be related to the
peculiar, sometimes alternating phase lags apparently 
associated with the 0.2 -- 10~Hz QPOs in GRS~1915+105
(\cite{cui99b,reig00,lin00}) and in XTE~J1550-564 
(\cite{wijnands99,cui00}). Assuming that the size scale
of the corona deduced from spectral fitting to GRS~1915+105
($\sim 10$ -- $100$~km) is correct, this would require that 
the inward-drift velocity of density perturbations is 
$\beta_{\rm r} \sim 10^{-4}$. It will be the subject of
future work to determine whether such a choice of $\beta_{\rm r}$
is realistic.

\section{\label{summary}Summary and conclusions}

A detailed re-analysis of the time-dependent radiation transfer
through hot, Comptonizing coronae in various models for small-scale
high-energy flares in X-ray binaries has been presented. In
contrast to previous work, the effect of coronal cooling due to
the increased soft photon input during the flare has been taken
into account self-consistently. In particular, the case of a slab
corona with a flare of the underlying, optically thick accretion
disc, the case of an inner, hot, quasi-spherical accretion flow
with a transition to an outer, cool, optically thick accretion
disc, and the case of the inward-drifting blob model have been
investigated.

In the slab-corona case, the resulting light curves look
drastically different from the ones calculated under the
assumption of a static corona. This had been pointed out
previously by Malzac \& Jourdain (\markcite{mj00}2000), who
had used a similar model setup as used in the slab-corona-case
section of this paper for two specific test cases, but did not
investigate the consequences of their results for the phase and
time lag spectra. In this paper, I have presented a systematic
study of the temporal features resulting from such a scenario.
The resulting phase and time lag spectra seem still quite well 
suited to explain many of the features observed, e.g., in 
Cyg~X-1. In particular, the overall shape of the time lag 
spectra, the photon-energy dependence of the time lags, and 
the small peak misalignment, consistent with 0, between the 
light curves at different photon energy bands, can be 
reproduced in such a geometry.

The model of a flaring disc surrounding a central hot, quasi-spherical
two-temperature flow, can be ruled out for most objects in which
significant phase and time lags have been observed since it would 
predict very small time lags (of the order of the light-crossing
time through the corona, i.e. $\lesssim 0.1$~ms), and light-curve
peak misalignments of the order of the maximum time lag, which is
generally not observed.

The light curves predicted by the inward-drifting blob model look
quite similar to the ones calculated under the assumption of a
static corona. However, there are marked differences with respect
to the details of the predicted phase and time lag spectra. Most
remarkably, the model (including coronal cooling) predicts oscillating
phase and time lag features with a typical frequency corresponding
to the inverse of the inward-drifting time scale. If these features
are to be associated with the alternating phase lags observed during
some observations of GRS~1915+105 and XTE~J1550-564 and apparently
associated with the 0.5 -- 10~Hz QPOs, this would require an inward
drift velocity of the density perturbations of $\beta_{\rm r} \sim
10^{-4}$.

The alternative model by Poutanen \& Fabian (\markcite{pf99}1999)
--- where the spectral variability is produced by a flare of the 
coronal heat input --- has been very successful in explaining the 
time-averaged photon spectrum and power, phase lag and coherence 
spectra of Cyg~X-1, and was in agreement with the peak of the 
cross-correlation functions between light curves at different 
energy bands being consistent with 0. The fundamentally different
flaring scenarios --- variation of the soft photon input vs. variation of
the coronal energy input --- should be distinguishable by virtue of their
markedly different spectral evolution and the photon-energy dependence
of the rms variability. In the case of a varying soft photon input, there
is considerable variability at soft X-ray energies, and the spectral
evolution resembles a pivoting of the hard X-ray spectrum around a
typical energy of $\sim 10$ -- 20~keV. Thus, up to an energy of $\sim
20$~keV, the fractional rms variability should decline as a function of
photon energy. In contrast, if the variability is powered by a varying
energy input into the corona, the soft X-ray band is expected to show
little variability, and the rms variability should increase with increasing
photon energy. Most interestingly, Cyg~X-1 shows both types of behavior
in different observations (\cite{lin00}). A consistently declining
rms variability over the {\it RXTE} PCA energy range has been observed
in GX~339-4, while 1E~1740.7-2942 almost always shows an increasing
rms variability with increasing photon energy (\cite{lin00}). This may 
indicate that fundamentally different variability processes are at work 
in different objects and even within the same object at different times.

In its original version, the Poutanen \& Fabian (\markcite{pf99}1999) 
model exclusively predicts positive phase lags, i.e. hard lags. The
negative and even alternating phase lags seen in GRS~1915+105 and
XTE~J1550-564 seem to indicate that those lags are not produced by
the variability of active coronal regions.

\acknowledgements{I thank the referee for useful comments, which
helped to improve, in particular, the discussion. The work of M.B. 
is supported by NASA through Chandra Postdoctoral Fellowship grant 
award no. 9-10007, issued by the Chandra X-ray Center, which is 
operated by the Smithsonian Astrophysical Observatory for and on
behalf of NASA under contract NAS~8-39073.}

\begin{deluxetable}{ccccc}
\tablewidth{14cm}
\tablecaption{Lightcurve fit parameters for the coronal response to
accretion disc flares in slab-coronal geometry for different flare
durations $\Delta t_{\rm flare}$. The coronal height is $h = 10^8$~cm,
corresponding to $h/c \approx 3 \times 10^{-3}$~s. Standard coronal
parameters are $\tau_{\rm T} = 1$, $kT_p = 100$~MeV. For definition
of the parameters $\tau_{\rm r}$, $\tau_{\rm d}$, and $t_0$ see 
Eq. (\ref{adc_lightcurves}).}
\tablehead{
\colhead{$\Delta t_{\rm flare}$ [s]} & 
\colhead{energy channel [keV]} & 
\colhead{$\tau_{\rm r}$ [$10^{-2}$ s]} & 
\colhead{$\tau_{\rm d}$ [$10^{-2}$ s]} &
\colhead{$t_0$ [$10^{-2}$ s]}}
\startdata
$10^{-3}$          & 3 -- 10       & 0.56 & 0.38 & 0.50 \nl
$10^{-3}$          & 10 -- 50      & 6.00 & 4.99 & 1.28 \nl
$10^{-3}$          & 50 -- 500     & 10.6 & 1.37 & 1.30 \nl
\hline
$3 \times 10^{-3}$ & 3 -- 10       & 0.95 & 0.38 & 0.59 \nl
$3 \times 10^{-3}$ & 10 -- 50      & 4.80 & 1.31 & 1.12 \nl
$3 \times 10^{-3}$ & 50 -- 500     & 2.89 & 1.16 & 2.18 \nl
\hline
$10^{-2}$          & 3 -- 10       & 2.05 & 7.48 & 1.38 \nl
$10^{-2}$          & 10 -- 50      & 1.77 & 1.04 & 2.04 \nl
$10^{-2}$          & 50 -- 500     & 1.25 & 0.87 & 3.07 \nl
\hline
\enddata
\label{adc_table}
\end{deluxetable}

\begin{deluxetable}{cccc}
\tablewidth{10cm}
\tablecaption{Lightcurve fit parameters for the inward-drifting
blob model. Coronal parameters: $\tau_{\rm T} = 0.75$, $n_p \propto 
r^{-3/2}$, virial proton temperature. Disk parameters: $r_{\rm in} 
= 60$~km, $kT_{\rm BB} (r_{\rm in}) = 1.0$~keV. $T_{\rm blob} \propto 
r^{-p}$, $\beta_{\rm r} = 0.05$. For definition of the parameters 
$\sigma$ and $\alpha$ see Eq. (\ref{blob_lc}).}
\tablehead{
\colhead{$p$} & 
\colhead{energy channel [keV]} & 
\colhead{$\sigma$ [ms]} & 
\colhead{$\alpha$}}
\startdata
0    & 1 -- 3       & ---  & 0.00 \nl
0    & 3 -- 10      & 37.7 & 0.81 \nl
0    & 10 -- 50     & 5.69 & 1.66 \nl
\hline
0.25 & 1 -- 3       & 9.90 & 0.20 \nl
0.25 & 3 -- 10      & 4.33 & 1.28 \nl
0.25 & 10 -- 50     & 3.16 & 2.09 \nl
\hline
0.5  & 3 -- 10      & 4.41 & 1.11 \nl
0.5  & 10 -- 50     & 2.62 & 2.05 \nl
0.5  & 50 -- 500    & 2.48 & 2.80 \nl
\hline
0.75 & 1 -- 3       & 3.26 & 1.59 \nl
0.75 & 3 -- 10      & 2.10 & 2.02 \nl
0.75 & 10 -- 50     & 1.91 & 2.59 \nl
\hline
\enddata
\label{blob_table}
\end{deluxetable}

\end{document}